\documentclass{aastex7}

\begin{document}

\title{A Refined Orbital Ephemeris for the Intermediate Polar FO~Aquarii and Evidence for a Period Derivative}

\author[orcid=0000-0003-4069-2817,gname='Peter', sname='Garnavich']{Peter Garnavich} 
\affiliation{Department of Physics and Astronomy, University of Notre Dame}
\email{pgarnavi@nd.edu}

\author[0009-0003-0467-4440]{Jordan Tweddale}
\affiliation{Department of Physics and Astronomy,
University of Notre Dame}
\email{jtweddal@nd.edu}

\author[0000-0001-7746-5795]{Colin Littlefield}
\affiliation{Bay Area Environmental Research Institute, Moffett Field, CA 94035 USA}
\email{clittlef@alumni.nd.edu}


\begin{abstract}

We combine more than 40 years of eclipse timings of the intermediate polar FO~Aqr and derive new orbital ephemerides. The eclipse times are best fit with a second order polynomial, implying an $8\sigma$ detection of a period derivative. The results show that the orbital period of FO~Aqr is currently increasing, with a derivative of $\dot{P}=6.65\pm0.81\times10^{-11}$. 

\end{abstract}
 

\section{Introduction} 

Intermediate polars (IP) are cataclysmic variables (CV) containing a moderately magnetic white dwarf (WD). The field strength of the WD is sufficiently low that its spin in not locked to the binary orbital period and a partial accretion disk can exist. FO~Aqr is a bright, well-studied IP with a secondary star that partially eclipses its disk every 4.85~hr. The time of eclipse is difficult to measure precisely because the WD generates a large amplitude spin modulation with a period of 20.9~minutes \citep{patterson98}. Here, we combine eclipse times spanning more than 40~years to improve the orbital ephemeris and search for evolution of the period.

\citet{patterson98} published a compilation of 54 eclipse times\footnote{Times for the \citet{patterson98} eclipses are Heliocentric Juian Day (HJD), while all other times have been converted to Barycentric Julian Day (BJD).} over 16 years that yielded an orbital period of 0.2020596(1)~days\footnote{The value within the parentheses indicates the uncertainty on the final digit.}. \citet{marsh96} derived a similar orbital ephemeris for FO~Aqr. Quoting from \citet{patterson98}, eclipse times ``were estimated by visual inspection of light curves after removal of the strong 21 minute signal". This is essentially the same method used to estimate eclipse times for the observations described below.

Between 2022 and 2025, we obtained fast cadence light curves of FO~Aqr with the 0.8-m Sarah L. Krizmanich Telescope (SLKT) on the University of Notre Dame campus. Eighteen of the light curves contained at least one eclipse and an example is displayed in Figure~1. To reduce the impact of the spin modulation on the time of eclipse, we applied the LombScargle package \citep[L-S;][]{lomb76,scargle82} implemented in astropy \citep{astropy22} to model the spin and beat modulation in each dataset. A model of the pulses was subtracted from each light curve before a polynomial was fitted to the eclipse and a time of minimum estimated. 

FO~Aqr was observed by Kepler/K2 in 2014 and by TESS in 2021 and 2025. We used the time of minimum published by \citet{kennedy16} based on the phased K2 light curve. For TESS, we constructed L-S models for each visit and subtracted them from the light curves. We then phased the residual light curves on the \citet{patterson98} orbital period and measured the averaged time of minimum referenced to the initial eclipse from each time-series. For the 2021 data, the averaged minimum was BJD~2459447.771(2), and for 2025 the averaged minimum was BJD~2460803.394(2). 

To fill in the gap between the 1990s and 2020s, we identified six light curves of FO~Aqr in the American Association of Variable Stars (AAVSO) International Database with sufficient cadence and data quality to allow good estimates of the eclipse times. The light curves were obtained by Berto Monard (AAVSO observer code MLF). The times of eclipse from the AAVSO light curves were estimated in the same manner as described in \citet{patterson98}.

FO~Aqr suffered a series of low-states beginning in 2016 \citep{littlefield20}. Compared to pre-2016 data, the light curves of FO~Aqr display lower spin pulse amplitudes over the second half of the binary orbit and narrower eclipse widths \citep{garnavich22}. The net impact has been to reduce the measurement uncertainties on eclipse times after recovery from the low states.

\begin{figure*}[t!]
\plotone{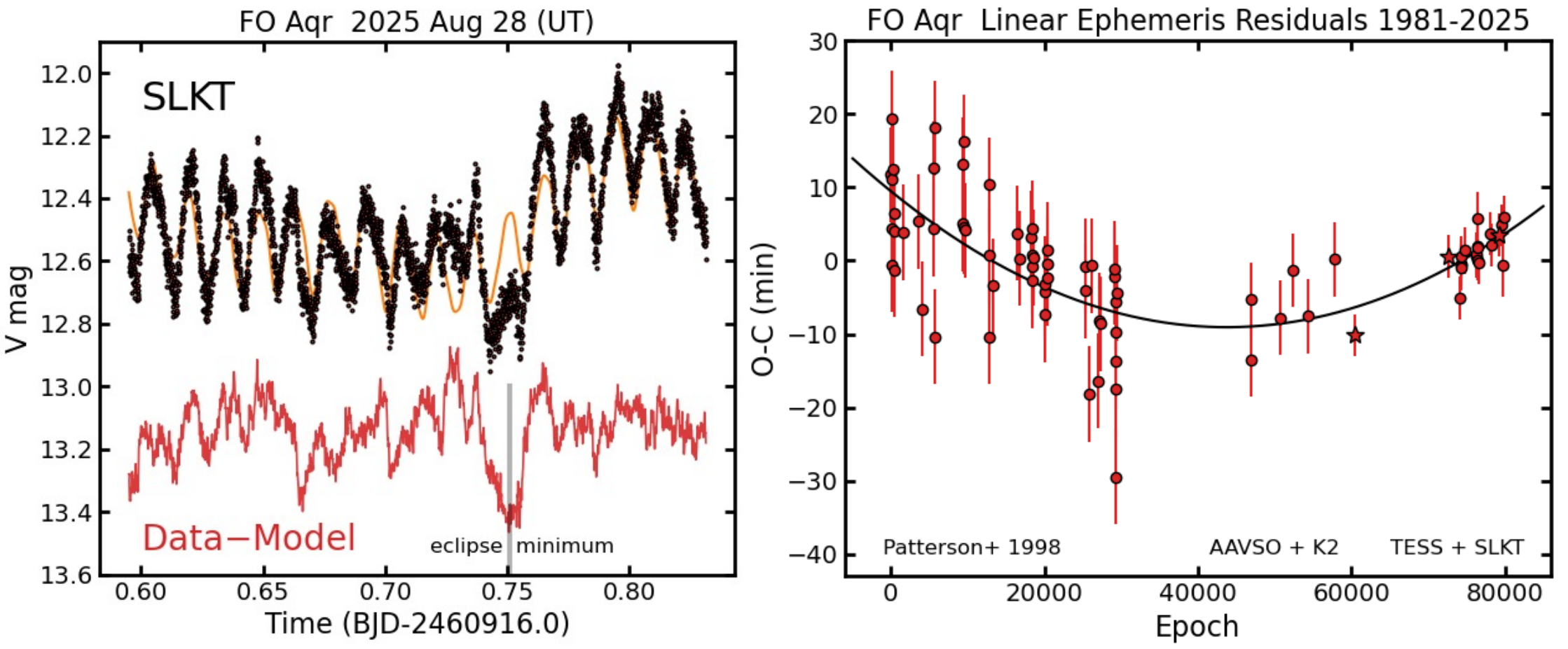}
\caption{{\bf Left:} The dark points show one of the recent light curves of a FO~Aqr obtained from the SLKT. This light curve covers approximately one binary orbit, while the rapid modulation results from the spinning white dwarf. The solid line plotted on the points is a light curve model constructed from L-S analysis. The difference between the data and model is plotted as a solid line below the light curve. The vertical line marks the measured time of the eclipse in the residual light curve.  {\bf Right:} The difference between the observed time of eclipse and the time calculated from the updated linear ephemeris in Equation~1. The K2 and TESS timings are shown with star symbols. The solid line displays a quadratic fit to the residuals that implies the orbital period of FO~Aqr is increasing.   
\label{figure1}}
\end{figure*}

\section{Results} 

Assuming the first eclipse listed by \citet{patterson98} as epoch zero, we calculated a weighted linear regression on the 81 eclipse times and find an ephemeris of:
\begin{equation}
T(BJD)=2444782.863(1)+E\; 0.20205999(2) \;\;\;\;,
\end{equation}
where $E$ is the epoch number and the values within parentheses are the uncertainties on the final digits. The $\chi^2$ parameter of this fit is 155 with 79 degrees of freedom (dof). The residuals to the fit are displayed in Figure~1 and show a curvature that suggests orbital period evolution. 

A weighted quadratic fit to the eclipse times results in the following ephemeris:
\begin{equation}
T(BJD)=2444782.869(1)+E\; 0.20205940(7)+E^2\; 6.72(81)\times10^{-12} \;\;\;\;.
\end{equation}
The reduced $\chi^2$ parameter from the quadratic fit is 83 for 78 dof, implying a significant improvement over the linear model. The coefficient of the quadratic term demonstrates the presence of a positive orbital period derivative where $\dot{P}= 6.65\pm0.81\times10^{-11}$, with the period increasing over a characteristic age of ${1\over2}P/\dot{P}\approx 4$~Myr.

\section{Discussion} 

Magnetic breaking is expected to drive CVs with orbits longer than 3~hrs to shorter periods \citep{rappaport83,knigge11}, and the mechanism provides a plausible explanation for the observed gap in the CV orbital period distribution.
\citet{schaefer24} presented orbital $\dot{P}$ estimates for a large number of CVs spanning many sub-classes. Both increasing (positive) and decreasing (negative) orbital periods were found with $\dot{P}$ amplitudes ranging over several orders of magnitude. The sign and amplitude we have found for FO~Aqr are not exceptional within the \citet{schaefer24} compilation. Few IPs were included in the study. The short period system EX~Hya was found to have a period derivative of $\dot{P}=-7.2\pm0.5\times10^{-13}$, but with significant deviations from the apparent long-term trend. DQ~Her does not yet show a detectable period evolution. \citet{knigge11} suggests that relatively short time-scale fluctuations in the donor star mass loss rate might explain the wide range of observed $\dot{P}$ values while the long-term average could match the standard model of CV evolution.

We thank B. Monard for excellent photometric observations of FO~Aqr posted to the AAVSO International Database. This work was partly supported by NASA Grant HST-GO-16727.

\facilities{TESS,SLKT,AAVSO}

All the {\it TESS} data used in this paper can be found in MAST: \dataset[10.17909/g203-sd46]{http://dx.doi.org/10.17909/g203-sd46}

\end{document}